\documentclass{aa}
\usepackage{graphicx,psfig,color}
\usepackage{multirow}
\usepackage{afterpage,lscape}
\usepackage[varg]{txfonts}
\usepackage{txfonts}
\newcommand{\microm}{\,\mu\rm{m}}

\newcommand{\degree} {^{\rm o}}

\newcommand{\AU}{\,\rm{AU}}

\def\mean#1{\left< #1 \right>}

\begin{document}

   \title{Testing dust trapping in the circumbinary disk around GG Tau A}

   \author{P. Cazzoletti         \inst{1,2,3}
\and       L. Ricci        \inst{2} 
\and       T. Birnstiel        \inst{2,4}
\and       G. Lodato        \inst{3} 
}

   \institute{Max-Planck-Institut f{\"u}r Extraterrestrische Physik, Gießenbachstraße, 85741 Garching bei München, Germany\\
              \email{pcazzoletti@mpe.mpg.de}
         \and
             Harvard-Smithsonan Center for Astrophysics, 60 Garden Street MS68, Cambridge, MA 02138, USA
          \and
             Universit{\'a} degli Studi di Milano, Via Giovanni Celoria 16, 20133 Milano, Italy
          \and
             Max-Planck-Institut f{\"u}r Astronomie, Königstuhl 17, 69117 Heidelberg, Germany
             }

   \date{Received XXX 2010/ Accepted YYY ZZZZ}

   \titlerunning{GG Tau Circumbinary Dust Ring Paper}

   \authorrunning{Cazzoletti et al.}
 
  %\abstract{[TO BE MODIFIED] The protoplanetary disk around GGTau A is so far the most studied circumbinary disk, and has been observed at many different wavelengths. Continuum mm/sub-mm observations detected a dust ring located between 200 AU and 300AU from the center of the system. Such a ring structure can be explained by the presence of a local maximum in the gas radial pressure profile which creates a “trap” for mm-sized dust grains, and likely induced by the dynamical interaction between the disk inner edge and the binary motion.

%We show that if the binary orbit and the disk are coplanar, tidal truncation of the circumbinary disk occurs at a radius which is too small compared to the inner edge inferred by the dust observation, and that in this “coplanar case” the pressure bump and the dust ring are located at around 100-150AU. On the other hand, we also show that in order for dust trapping to occur at the observed radius, the disk and orbital plane must be misaligned of ~27 degrees. In the “misaligned case” our simulations reproduce the mm-size dust ring. We run the simulations for different values of viscosity, and find that observations are reproduced with the greatest accuracy for alpha=0.002.}

 \abstract
  % context heading (optional)
  % {} leave it empty if necessary  
   {The protoplanetary disk around the GGTau A binary system is so far one of the most studied young circumbinary disk, and has been observed at many different wavelengths. Observations of the dust continuum emission at sub-mm/mm wavelengths detected a dust ring located between 200 AU and 300 AU from the center of  mass of the system. According to the classical theory of tidal interaction between a binary system and its circumbinary disk, the measured inner radius of the mm-sized dust ring is significantly larger than the predicted truncation radius, given the observed projected separation of the stars in the binary system (0.25'', corresponding to $\sim34\,\rm AU$). 
A possible explanation for this apparent tension between observations and theory is that a local maximum in the gas radial pressure is created at the location of the center of the dust ring in the disk as a result of the tidal interaction with the binary.
An alternative scenario invokes the presence of a misalignment between the disk and the stellar orbital planes. }
  % aims heading (mandatory)
   {We investigate the origin of this dust ring structure in the GG Tau A circumbinary disk, test whether the interaction between the binary and the disk can produce a gas pressure radial bump at the location of the observed ring, and discuss whether the alternative hypothesis of a misaligned disk offers a more viable solution.}
  % methods heading (mandatory)
   {We run a set of 3D hydrodynamical simulations for an orbit consistent with the astrometric solutions for the GG Tau A stellar proper motions, different disk temperature profiles, and for different levels of viscosity. Using the obtained gas surface density and radial velocity profiles, we then apply a dust evolution model in post-processing in order to to retrieve the expected distribution of mm-sized grains.}
  % results heading (mandatory)
   {Comparing the results of our models with the observational results, we show that, if the binary orbit and the disk were coplanar, not only the tidal truncation of the circumbinary disk would occur at a radius that is too small compared to the inner edge inferred by the dust observations, in agreement with classical theory of tidal truncation, but also that the pressure bump and the dust ring in the models would be located at $<$150~AU from the center of mass of the stellar system. This shows that the GG Tau A circumbinary disk cannot be coplanar with the orbital plane of the binary. We also discuss the viability of the misaligned disk scenario, suggesting that in order for dust trapping to occur at the observed radius, the disk and orbital plane must be misaligned by an angle of about 25-30 degrees.}
  % conclusions heading (optional), leave it empty if necessary 
   {}


   \maketitle
%
%________________________________________________________________

\section{Introduction}

Most stars form in multiple systems \citep[e.g.][]{1991A&A...248..485D,1992ApJ...396..178F,2010ApJS..190....1R}. It is also undisputed that the majority of stars in young star forming regions show direct or indirect evidence for the presence of a young circumstellar disk. Moreover, also disks orbiting the whole multiple stellar system, called \textit{circumbinary} in the case of binary systems, are sometimes observed \citep[e.g.][]{1995ApJ...450..824S}. Planets originate from these disks and so far ten circumbinary planets (sometimes called "Tatooine planets") have been discovered by Kepler orbiting around eight eclipsing binaries \citep{2011Sci...333.1602D,2012Natur.481..475W,2012ApJ...758...87O,2012Sci...337.1511O,2013ApJ...768..127S,
2013ApJ...770...52K,2014ApJ...784...14K,2014IAUS..293..125W}. Planets, therefore, are common in binary (or multiple) system.

Disk-binary interaction plays a pivotal role in such systems. In binaries, part of the material of a disk around one or both stars is ripped away by tidal forces. The net result of the interaction between the gas disk and the binary is a net exchange of angular momentum via tidal torques. In particular, the angular momentum of the disk-binary system is transported outwards \citep{lp79b,lp79a}. This means that gas in the individual circumstellar disks loses angular momentum to the binary and moves toward inner orbits; the gas in circumbinary disks, on the other hand, acquires angular momentum from the binary and is repelled from the central stars. Disk viscosity tends to contrast the effect of tidal torques. At a certain radius, called \textit{tidal truncation radius}, viscous and tidal torques balance each other and an equilibrium configuration is reached \citep{lp86b}.

Theoretical studies sought to find a model capable of providing an estimate of the radius of each one of the three disks in a binary system (i.e. circumprimary, circumsecondary and circumbinary), given the key orbital parameters (semi-major axis \textit{a}, eccentricity \textit{e}, mass ratio \textit{q} and inclination \textit{i} between disk and orbital plane) and some characterization of the disk viscosity \citep{pp77,pac77,al,pichardo,mirandalai,nixon}. In one of the most comprehensive works so far \cite{al} estimated the truncation radius for binaries coplanar with the disks and for any value of \textit{q}, \textit{e} and viscosity. \cite{mirandalai} recently generalized this study for misaligned systems. 

Observations of young disks in multiple systems have the potential to test the predictions of tidal truncation theory \citep{harrisetal}. The best cases are offered by the systems in which the proper motion for the stellar components are measured for a significant fraction of the stellar orbits. In this case, orbital solutions are provided from the analysis of the proper motions, and models incorporating the tidal interaction between the multiple system and the disks can be used to predict the spatial distribution of gas and dust in the disks. High angular resolution observations of the disks emission can then be used to test the predictions of those models. In this work we use the results of recent observations for the GG Tau A circumbinary disk to test models of tidal truncation. 

The young quadruple system GG Tau has been the subject of many different studies. It is composed of two low-mass binary systems, namely GG Tau A and GG Tau B. GG Tau Aa and GG Tau Ab, respectively of mass $0.78\pm0.09\, \rm M_{\odot}$ and $0.68\pm0.02\,\rm M_{\odot}$ 	\citep{white99} and with an angular separation of $0.25''$ \citep{lein93}, form together GG Tau A, one of the most studied and best known nearby \citep[$\sim140\,\rm pc$, ][]{1978ApJ...224..857E} young binary systems \citep[$\sim1\,\rm Myr$, ][]{2001ApJ...556..265W}\footnote{Note that VLTI/PIONIER and VLT/NACO observations suggest that GG Tau Ab is itself a close binary with two stars with very similar mass at a projected separation of $\approx 4$ AU \citep{2014A&A...565L...2D}. Given the small separation relative to the distance to the circumbinary disk, which is the subject of our study, we will consider GG Tau Ab1 and GG Tau Ab2 as a single component GG Tau Ab.}. Its circumbinary disk has been observed in both dust thermal emission \citep{dutrey94,guillo99,andrews14}, scattered light emission \citep{1996ApJ...463..326R,2000ApJ...536L..89S,duchene} and $\rm CO$ gas emission \citep{dutrey14}, and has a total mass of $\sim0.12\, M_{\odot}$ and an inclination $i=37\degree$ with respect to the line of sight \citep{guillo99} . The other binary system GG Tau B, located $10.1''$ south, is wider ($1.48''$) and less massive \citep{lein93}.

\begin{figure}[htbp!]
\begin{center}
\includegraphics[width=\columnwidth]{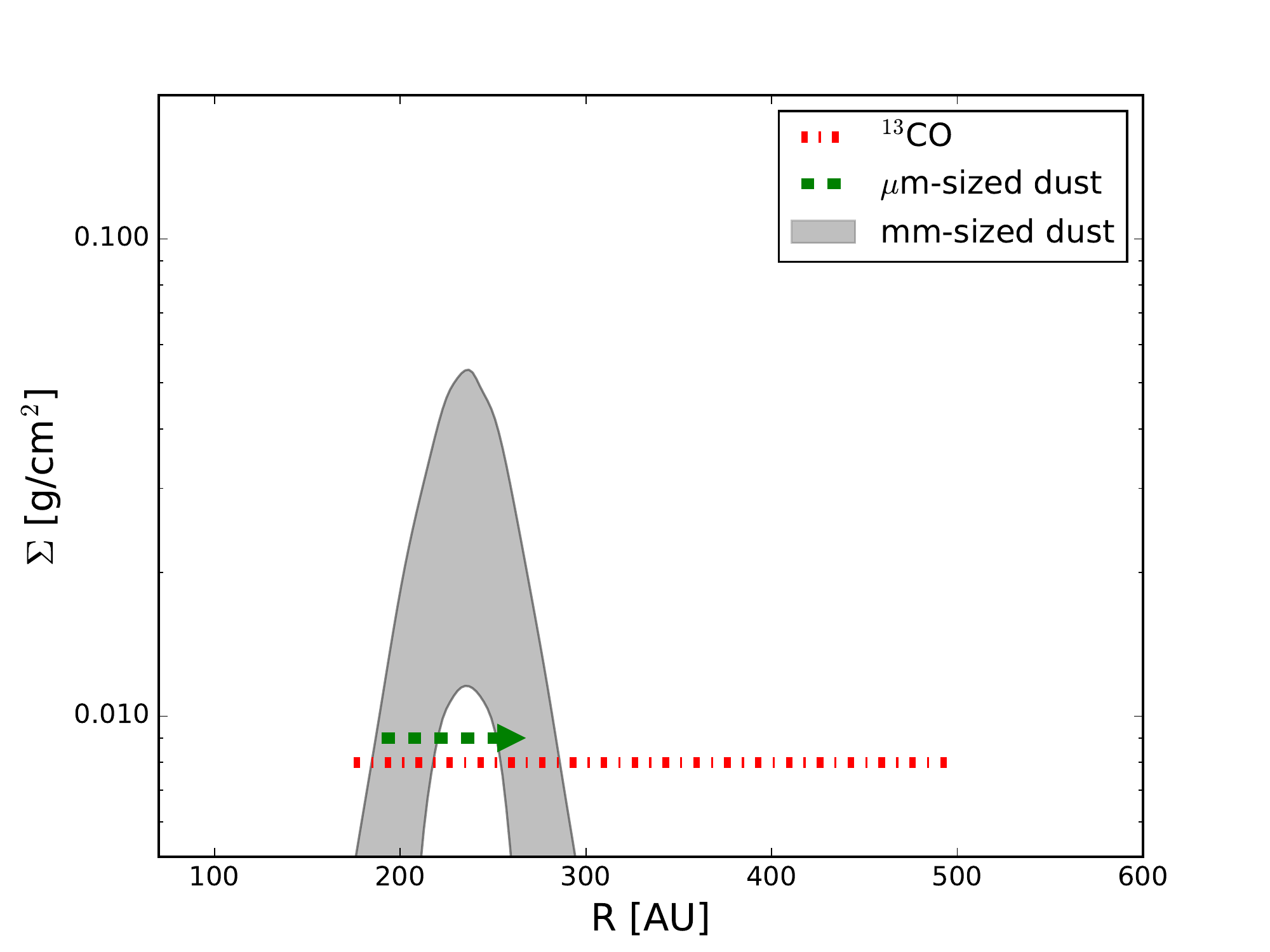}
\caption{Observational constraints available to date for different sized grains and for CO distribution. The grey-shaded area shows the $2\sigma$ uncertainty on the mm-sized dust radial density profile as modelled in \citet{andrews14}. The red dashed-dot and the green dashed lines show the radial extent of CO and micron-sized dust grains, respectively. It should be noted that these two tracer are optically thick, and do not provide information about the density profile. The vertical location of the lines is therefore arbitrary.}\label{fig:obs} 
\end{center}
\end{figure}

\begin{table*}[htbp!]
\caption{Parameters of the best orbital solutions. The first column is obtained by assuming the disk to be coplanar with the orbit, and therefore by fixing the value of $i$. The other ones have been obtained instead by fixing the value of a. The data on the first two columns is from \cite{kohler}, the third and the fourth are a personal communication of the author of the paper.}
\label{tabkoehler}
\begin{center}
\begin{tabular}{lr@{}lr@{}lr@{}lr@{}l}
\hline
\hline
Orbital Element			& \multicolumn{2}{c}{Orbit coplanar} & \multicolumn{2}{c}{$a=60\,\rm AU$} & \multicolumn{2}{c}{$a=70\,\rm AU$} & {$a=80\,\rm AU$} \\
				& \multicolumn{2}{c}{with disk}  & \multicolumn{2}{c}{constraint} & \multicolumn{2}{c}{constraint} & {constraint} \\
\noalign{\vskip1pt\hrule\vskip3pt}
Date of periastron $T_0$ 	& $2477680$ & $\,^{+690}_{-270}$  & \quad$2463400$ & $\,^{+1470}_{-5420}$ & $2462334$ & $\,^{+1436}_{-5222}$ & $2461114$ & $\,^{+2013}_{*****}$ \\[3pt]
(Julian)			& (July 2071)\span		 & (June 2032)\span		& (2029 Jul 16)\span & (2026 Mar 14)\span \\[2pt]
Period $P$ (years)		& $  162  $ & $\,^{+62}_{-15}$	 & $ 403 $ & $\,^{+67}_{-32}$	& $ 507$ & $\,^{+83}_{-41}$ & $ 638$ & $\,^{+77}_{-68}$\\[3pt]
Semi-major axis $a$ (mas)	& $  243  $ & $\,^{+38}_{-10}$	 & $ 429 $	&  & $  500.0$ & & $  571.4$ \\[3pt]
Semi-major axis $a$ (AU)	& $   34  $ & $\,^{+5.9}_{-2.8}$  & $  60 $	&  & $  70.00$ & & $  80.00$ \\[3pt]
Eccentricity $e$		& $   0.28$ & $\,^{+0.05}_{-0.14}$& $  0.44$ & $\,^{+0.02}_{-0.03}$ & $  0.510$ & $\,^{+0.017}_{-0.013}$ & $  0.565$ & $\,^{+0.004}_{-0.007}$ \\[3pt]
Argument of periastron
	$\omega$ ($^\circ$)	& $  91   $ & $\,^{+4}_{-13}$	 & $  19  $ & $\,^{+9}_{-10}$	& $-165$ & $\,^{+ 8}_{-13}$ & $-169$ & $\,^{+ 10}_{ -2}$ \\[3pt]
P.A. of ascending node
	$\Omega$ ($^\circ$)	& $ 277   $ & $\,^{+2.0}_{-2.0}$  & $ 131  $ & $\,^{+13}_{ -8}$	& $ 133$ & $\,^{+11}_{ -7}$ & $ 135$ & $\,^{+11}_{ -8}$ \\[3pt]
Inclination $i$ ($^\circ$)	& $ 143   $ & $\,^{+1.3}_{-1.0}$  & $ 132.5$ & $\,^{+1.0}_{-2.5}$	& $ 131.1$ & $\,^{+ 0.9}_{ -0.6}$ & $ 131.4$ & $\,^{+ 0.9}_{ -0.6}$ \\[3pt]
Angle between orbit and disk	& $   0.02$ & $\,\pm1.9$	 & $  24.9$ & $\,\pm1.7$	& $	26.9$ &  &$	28.1$ \\
\hline
\hline
\end{tabular}
\end{center}
\end{table*}

The disk around GG Tau A shows a peculiar ring-shaped dust distribution.
\cite{andrews14} modelled the continuum dust emission measured at several millimeter wavelengths with a radial distribution for mm-sized grains as a gaussian peaked at $235\pm 5\,\rm AU$, and with a very narrow width, $FWHM\sim60\,\rm AU$. Imaging in scattered light at near infrared wavelengths have shown that the inner radius of the disk in smaller, so that $\mu$m-sized particles lie between $180$ and $190\,\rm AU$ \citep{duchene}. ALMA observations of the rotational transition $J= 3 - 2$ of $^{13}$CO show an intensity radial profile with a peak around $175-180$ AU from the binary center \citep{2016ApJ...820...19T}. However the relatively poor spatial resolution of those observations ($\approx 50$ AU) does not allow one to infer with precision the location of the inner radius in gas. Emission from both $^{12}$CO$(J=3-2)$ and $^{13}$CO$(J=3-2)$ is detected up to $\sim 500$ AU from the binary, clearly indicating that gas is found also at much larger radii than the mm-sized dust. Among all the components observed, only mm-sized dust grains are optically thin and can therefore provide information on the density profile of mm dust. For this reason, we will mainly focus our study on the dust ring observed at submillimeter wavelengths. A summary of the various constraints for the disk size in the various components is shown in Fig. \ref{fig:obs}.

In order to use this information on the spatial extent of the GG Tau A circumbinary disk to test the predictions of tidal truncation models, some constraints on the orbital parameters of the GG Tau A binary system are needed. For example, the classical calculations by \cite{al} predict disk truncation radii at 2 or 3 times the value of the binary semi major axis \textit{a}. 
A number of astrometric observations for the GG Tau A system are available, and span almost twenty years since 1990 to 2009. However, these measurements cover only a fraction of the orbital period of the binary and do not allow to constrain all the orbital parameters at the same time. Fixing one of the orbital parameters, all the others can be obtained by fitting the proper motions \citep{kohler}. Tab. \ref{tabkoehler} shows four different orbits consistent with the astrometric measurements: the first column is obtained by requiring that the disk and the binary orbit are coplanar (i.e. putting constraints on the inclination $i$ of the orbit and on the position angle of its ascending node), while the other three are calculated by fixing the value of the semi-major axis of the orbit to $60,70,$ and $80\,\rm AU$, respectively.

If the inclination \textit{i} and the position angle of the ascending node of the binary orbit are constrained by requiring the orbit to be coplanar with the disk, then $a \approx 34$ AU. In this case the dust inner ring is located much further out than the predicted $\sim100$ AU gas inner truncation radius. A possible explanation for this apparent discrepancy was proposed by \cite{andrews14}: if the gas radial density profile at the inner edge of the disk were very shallow, then the pressure maximum where mm-size grains drift toward and accumulate would lie at a much larger radial position than the gaseous disk inner radius. 
Hydrodynamical simulations calculating the expected gas radial profile are needed to test this hypothesis. 

Alternatively, if one relaxes the hypothesis of coplanarity, the astrometric measurements can be fit by orbits misaligned with respect to the disk plane and with larger values of \textit{a} (as shown in Tab. \ref{tabkoehler}). These misaligned configurations would allow the binary system to dynamically truncate the disk at a location closer to the observed dust ring. In this misaligned case, also a steeper radial profile for the gas density would in principle be able to produce a dust ring at radii even larger than 200 AU.

It should be noted that a coplanar disk-binary system with $a>34\,\rm AU$ is not entirely ruled out by the astrometric measurements, and that a coplanar orbit with a larger semi-major axis is still possible within a $5\sigma$ uncertainty. Due to the much lower likelihood of this solution, however, we decide not to address it in this work. More astrometric measurements are needed for a more detailed analysis.

In this paper we run a set of hydrodynamical simulations that account for tidal interaction between the binary and the circumbinary disk to calculate the predicted distribution of gas density in the best-fit case of coplanar disk and binary system. We couple a dust evolution model \citep{birnstiel} to the results of the hydrodynamical simulations for the gas to obtain predictions for the radial profile of the dust density.
We compare these predictions with the main features observed for the dust in the disk in order to test the validity of the coplanar hypothesis. We also discuss whether the alternative hypothesis of a misaligned disk could be a likely explanation for the location of the dust observed in the GG Tau A circumbinary disk. For this study, we always consider orbital parameters that are consistent with the measured stellar proper motions.

In Section \ref{secsimul} we briefly present the setup of our simulations. In Sections \ref{secresultsggtau} and \ref{secdiscggtau} we show the results we obtained in our simulations and we discuss the implication of our results have in the the search for an explanation for the observed narrow mm-dust ring.

\section{Methods}\label{secsimul}
\subsection{Gas simulations}\label{sec:sph}
In our work we use the \texttt{PHANTOM} Smoothed Particle Hydrodynamics (SPH) code \citep{lodp2010,2010MNRAS.406.1659P} in order to perform 3D hydrodynamical simulations for the gas alone.  The code computes the viscous evolution of a gas distribution in a disk by solving the equations of hydrodynamics in the presence of a gravitational field generated by one or two central stars, and/or a planetary mass companion.  For our purposes, we use a circumbinary disk of which we neglect the self-gravity. We want to study the resulting gas radial density and velocity profiles.

To mimic disk viscosity, we adopt the formulation by \citet{1994ApJ...431..754F}, where the stress tensor in evaluated directly in the Navier-Stokes equations.We can express the shear viscosity using the \citet{ss} prescription
\begin{equation}
\nu=\alpha\frac{c_{\rm s}^2(R)}{\Omega(R)},
\end{equation}
where $\alpha$ is the chosen value for the  \citet{ss} parameter, and $c_{\rm s}(R)$ and $\Omega(R)$ are the sound-speed and angular velocity radial profiles, respectively. The accuracy of this formulation has been tested also for physical phenomena strongly dependent on the chosen value of $\alpha$, such as the dynamics of warps \citep{lodp2010,2013MNRAS.433.2142F}.

Since the disk viscosity strongly affects both the location of the tidal truncation radius and the dust dynamics, we choose to run a set of different simulations using three different values of viscosity, corresponding to $\alpha$ of 0.01, 0.005 and 0.002. 

A second source of viscosity is also present. SPH codes implement an artificial viscosity in order to be able to resolve discontinuities by spreading them over a few smoothing lengths and to prevent particle interpenetration. This artificial term can be understood as a numerical representation of second derivatives of the velocity \citep{al,1996MNRAS.279..402M,lodp2010}; the resulting artificial viscosity parameter is given by
\begin{equation}\label{eq:artvisc}
\alpha_{\rm art}\approx\frac{1}{10}\alpha^{\rm AV}\frac{\mean h}{H},
\end{equation}
where $\mean h$ is the azimuthally averaged smoothing length (which is proportional to $n^{-1/3}$, $n$ being the local density of SPH particles), \textit{H} is the disk thickness, and $\alpha^{\rm AV}$ is set to a minumum value $\alpha^{\rm AV}_{\rm min}$ and increases up to a value $\alpha^{\rm AV}_{\rm max}$ in the presence of shocks by means of a \cite{1997JCoPh.136...41M} switch.  Note also that we used the notation $\alpha_{\rm art}$ to discriminate between the physical viscosity due to the artificial one and the directly implemented physical viscosity, for which we used $\alpha$. The total viscosity in SPH is therefore given by $\alpha_{\rm tot}\approx\alpha+\alpha_{\rm art}$. 

Since $\mean h\propto n^{-1/3}$, we can make the contribution of the artificial viscosity to the physical viscosity negligible by increasing the number of particles in the low-viscosity simulations (see Section \ref{sec:initcond}), thus making $\alpha_{\rm tot}\sim\alpha$.

In the simulations presented in this work, we set $\alpha^{\rm AV}_{\rm min}=0.03$ and $\alpha^{\rm AV}_{\rm max}=0.1$.  The \cite{1950JAP....21..232V} $\beta^{\rm AV}$ parameter was set equal to 2. We let each simulation evolve over 1000 binary orbital periods in order to reach steady state. We then compute the gas density and the gas radial velocity profile by averaging the quantities azimuthally. We also average them on a few orbital periods, in order to smooth the profiles and to remove the fluctuations due to the discretization of the fluid operated by SPH.

\subsection{Dust Simulations}
We use the gas density and radial velocity profiles obtained from our SPH simulations as a stationary "environment" where we let dust evolve following the model from \citet{birnstiel}.  We assume that the gas density in a binary system reaches a stationary state on time scales faster than the dust evolution time scales, which for typical dust-to-gas ratios are of the order of hundreds of local orbits (e.g. Brauer et al. 2008) that at the position of the dust ring is long compared to the binary orbital time scale ($\sim 200 \,\rm years$). We use this stationary gas density distribution as input for a global model of dust evolution (Birnstiel et al. 2010) to test how dust evolves under the physical conditions predicted by the SPH hydrodynamical simulations.

Our dust evolution model accounts for compact dust growth, cratering and fragmentation, radial drift, turbulent mixing and gas drag. In order to calculate the relative velocity of the dust particles, Brownian motion, turbulence, vertical settling, radial and azimuthal drift are taken into account. In these simulations, the initial size of all the particles is assumed to be $\sim 1\microm$. At the beginning of the growth process, when particles have still sizes of a few microns,  the main contribution to their relative velocities comes from Brownian motion and settling. In these early stages, growth by coagulation is very efficient and is a result of van der Waal's interaction between small grains. As they grow to larger sizes, they start to decouple from the gas, and turbulence as well as radial drift become the main sources of their relative velocities.

As the grains grow, their relative velocities increase \citep{birnstiel}. When dust grains reach sizes with high enough velocities, collisions no longer produce coagulation only, but dust also encounters destructive collisions. The threshold velocities above which fragmentation becomes dominant can be estimated through laboratory experiments and theoretical work of collisions for silicates and ices \citep[e.g.][]{2008ARA&A..46...21B,2007A&A...470..733S,2009ApJ...702.1490W}. For the silica particles these threshold velocities are of the order of a m/s, and they increase with the presence of ices \citep{2015ApJ...798...34G}. We ran some test models with different values of fragmentation velocity $v_{\rm frag}$ and we found that the results were not significantly affected. In our models we adopted a standard value of $v_{\rm frag}=10\,\rm m/s$.

The level of coupling between the dust and the gas is quantified with the dimensionless stopping time $\tau_{\rm fric}$, defined as the ratio between the stopping time of the particle due to friction with the gas and the orbital time-scale $\Omega$. Particles with $\tau_{\rm fric}\gg1$ are decoupled from the gas, they are not affected by any drag force and therefore rotate around the star on their own Keplerian orbit. On the other hand, particles with $\tau_{\rm fric}\ll1$ are strongly coupled with the gas, and move along with it. Particles experiencing the biggest radial drift are the ones characterized  by $\tau_{\rm fric}=1$. In the case of GG Tau A disk, in the vicinity of the observed dust ring we expect this to occur for grains with sizes of $\sim1-10$ mm. Due to the sub-Keplerian rotation velocity of the gas, these particles experience a gas headwind that leads them to lose angular momentum and to drift radially towards the disk inner regions \citep{1972fpp..conf..211W,1986Icar...67..375N,2007A&A...469.1169B}.

One of the biggest unknowns for dust evolution models is whether dust growth is compact or fractal. In our models we assume compact growth. However, fractal and compact growth models are not expected to produce significantly different results in terms of the sub-mm emission from the disk outer regions. This is because these different modes of solid growth produce particles with similar $\tau_{\rm fric}$ (even though they have different size and filling factor) and absorption/emission dust opacities are proportional to $\tau_{\rm fric}$. Therefore, we do not expect this intrinsic uncertainty of the models to play an important role on the results of the work presented here.
    
In section~\ref{sec:dust_coplanar} we apply this dust model to simulate the behavior of dust particles in the coplanar case for the GG Tau A circumbinary disk. We use these results to compare the predictions of our models to the radial distribution of dust particles as constrained by the observations.    
    
\subsection{Initial Conditions}\label{sec:initcond}
We tune our initial conditions to reproduce the main characteristics of the GG Tau A system. First of all the eccentricity \textit{e} and the semimajor axis \textit{a} of the orbit are set accordingly to the best-fit orbits calculated by \cite{kohler} to reproduce the measured stellar proper motions (Tab. \ref{tabkoehler}). In this work we are interested in simulating in detail the case of coplanar disk and binary orbital plane (second column in Tab. \ref{tabkoehler}), but we will also discuss the hypothesis of a disk misaligned with the binary which allow for larger values of the semi-major axis than in the coplanar case (a few possible cases are listed on the third, fourth and fifth column in Tab. \ref{tabkoehler}). 

In the SPH simulations, the two stars are modelled as sink particles \citep{1995MNRAS.277..362B}  with mass $0.78\, M_{\odot}$ and $0.68\, M_{\odot}$, respectively \citep{white99}. Each of the two sink particles has an associated accretion radius, i.e. a radius within which we can consider gas particles to be accreted onto the stars. Since for our purposes we do not need to know what happens to the gas in the vicinity of the stars, we can set the sink radii to fairly large values, thus speeding up the simulations. In particular, we use $R_{\rm sink}=0.1\,a$. 

We set the initial disk inner and outer radii at $t=0$ to $R_{\rm in}=2a$ and  $R_{\rm out}=800\AU$, respectively. Between these two edges, the initial gas density profile we use is
\begin{equation}\label{eq:dens}
\Sigma(R)=\Sigma_0\frac{a}{R}\Bigg(1-\sqrt{\frac{R_{\rm in}}{R}}\Bigg),
\end{equation}
where $\Sigma_0$ is a normalisation factor. Its value is chosen in each simulations in order to give a total gas mass of around  $\sim0.12M_{\odot}$ \citep[the disk mass estimated by ][]{guillo99}. We then set a Keplerian velocity profile for the gas, relative to a $1.46\, M_{\odot}$ central mass. It should be noted that our hydrodynamical results do not depend on the initial gas density profile, since we will let our system evolve until steady state is reached. We assume for the gas a locally isothermal equation of state, where the temperature along the \textit{z} axis at each radius is fixed. To describe the temperature radial profile we adopted the one inferred from $^{13}\rm CO$ measurements and analysis by \citet{guillo99}:
\begin{equation}\label{eq:tprof}
T(R)=20\,\rm K\bigg(\frac{R}{300\,\rm AU}\bigg)^{-0.9}.
\end{equation}
In order to test the effect of this choice for the disk temperature on the results of our study, we also ran simulations with a less steep temperature radial profile ($T\propto R^{-0.5}$, see Appendix A). 

The temperature profile is related to the disk thickness \textit{H} by assuming vertical hydrostatic equilibrium:
\begin{equation}\label{eq:hr}
{H(R)}=\frac{c_{\rm s}(R)}{\Omega_{\rm K}}=\sqrt{\frac{k_{\rm B}T(R)}{\mu m_{\rm p}}}\frac{1}{\Omega_{\rm K}},
\end{equation}
where $\mu=2.3$ is the mean molecular weight and $k_{\rm B}$ is the Boltzmann constant. The particles are then distributed in the vertical direction so to obtain a Gaussian density profile with thickness $H(R)$.  Combining Eq. \ref{eq:tprof} and Eq. \ref{eq:hr} we get $H/R\approx0.12$ in our simulations.

Among all the factors that affect the evolution of gas and dust in the disk, viscosity plays an important role. For this reason we decided to run different simulations for different values of viscosity.  When simulating disks with lower viscosities, we correspondingly increased the number of SPH particles to reduce possible effects given by the artificial viscosity, as explained in section~\ref{sec:sph}. In particular we ran simulations with $\alpha=0.01$ (using $10^6$ particles), $\alpha=0.005$ (using $3\times10^6$ particles), and $\alpha=0.002$ ($6\times10^6$ particles). We obtain $\alpha_{\rm art}\leq0.002$ in the $10^6$ particles simulation, $\alpha_{\rm art}\leq0.0015$ in the $3\times10^6$ one, and $\alpha_{\rm art}\leq0.001$ when using  $6\times10^6$ particles.

\begin{figure}[htbp!]
\begin{center}
\includegraphics[width=\columnwidth]{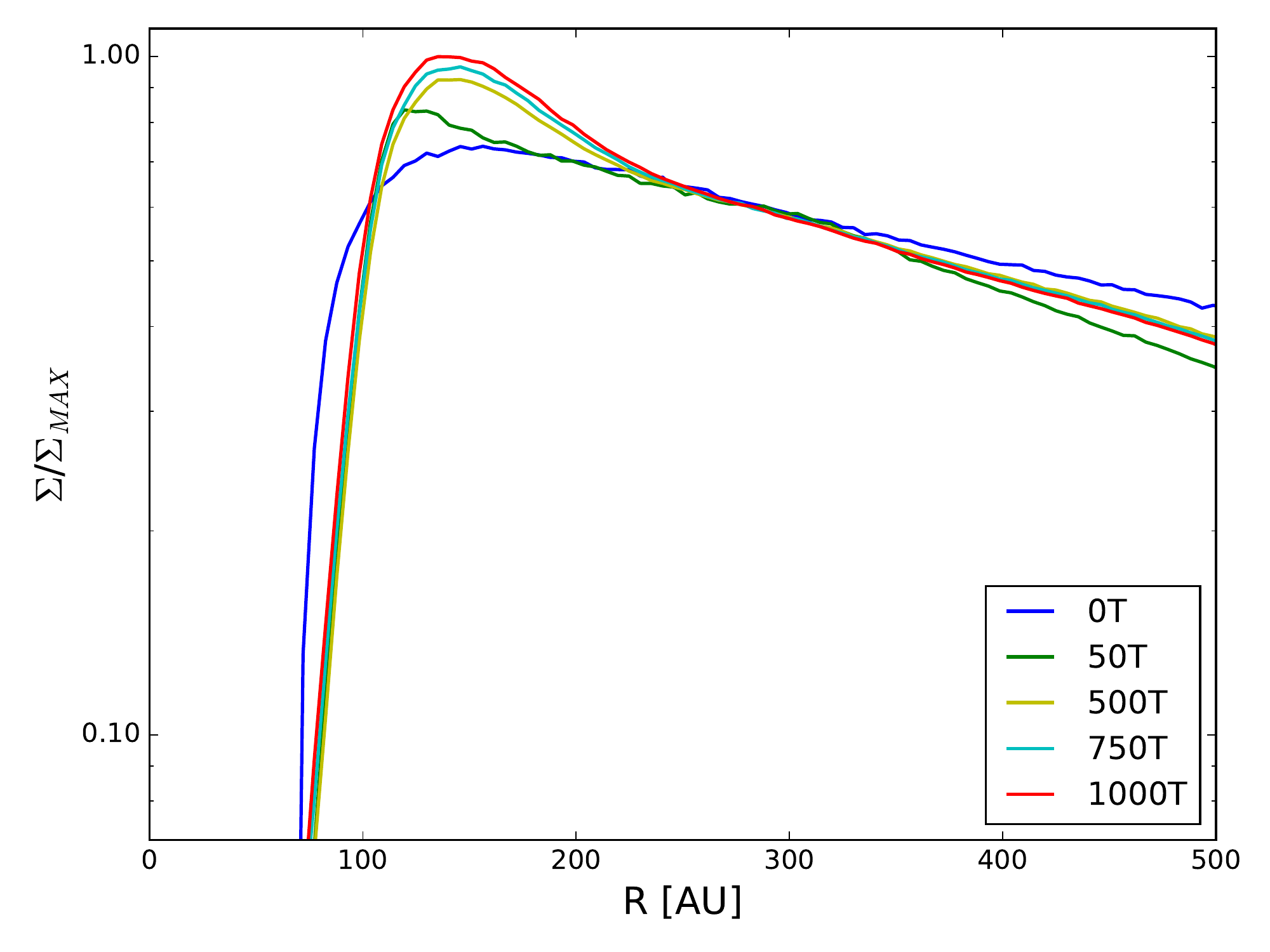}
\caption{Azimuthally averaged gas density profile for the $\alpha_{\SS}=0.002$ simulation at different evolutionary stages. After few hundreds binary orbits the density profile reaches a quasi-stationary configuration. All the density profiles are normalized to the maximum of the density profile at the end of the simulation.}\label{fig:timeev} 
\end{center}
\end{figure}

\section{Results}\label{secresultsggtau} 
In this section we present the results obtained with our simulations with a binary orbiting on the same plane of the disk. In this case, the binary orbit has $a=34\AU$, $e=0.28$. We study how the initial gas density profile evolves with time, and how it is affected by the chosen values of viscosity. Particular focus is given to how time and viscosity affect the location and shape of the gas density maximum, because marginally coupled dust particles will radially drift. to this region. 
In these simulations, the binary orbit has semi-major axis and eccentricity of $a=34\AU$ and $e=0.28$, respectively, corresponding to the best-fit orbit solutions of the measured stellar proper motions (second column in Table~\ref{tabkoehler}).  
\subsection{Time evolution of the gas density profile}
We first verify that the time we let our simulation evolve for (1000 orbital periods) is sufficient for the gas density profile to reach a steady state configuration. Fig. \ref{fig:timeev} shows a comparison of the gas density profile at different times for the $\alpha=0.002$ simulation, i.e. the one with the longest viscous timescale. Even for the lowest viscosity, the density profiles after 750 and 1000 orbital periods differ by at most $\sim3$\%. We are therefore allowed to assume the disk to have reached steady state after 1000 orbital periods.

\subsection{Different viscosities}\label{sec:viscosities}
\begin{figure}[htbp!]
\begin{center}
\includegraphics[width=\columnwidth]{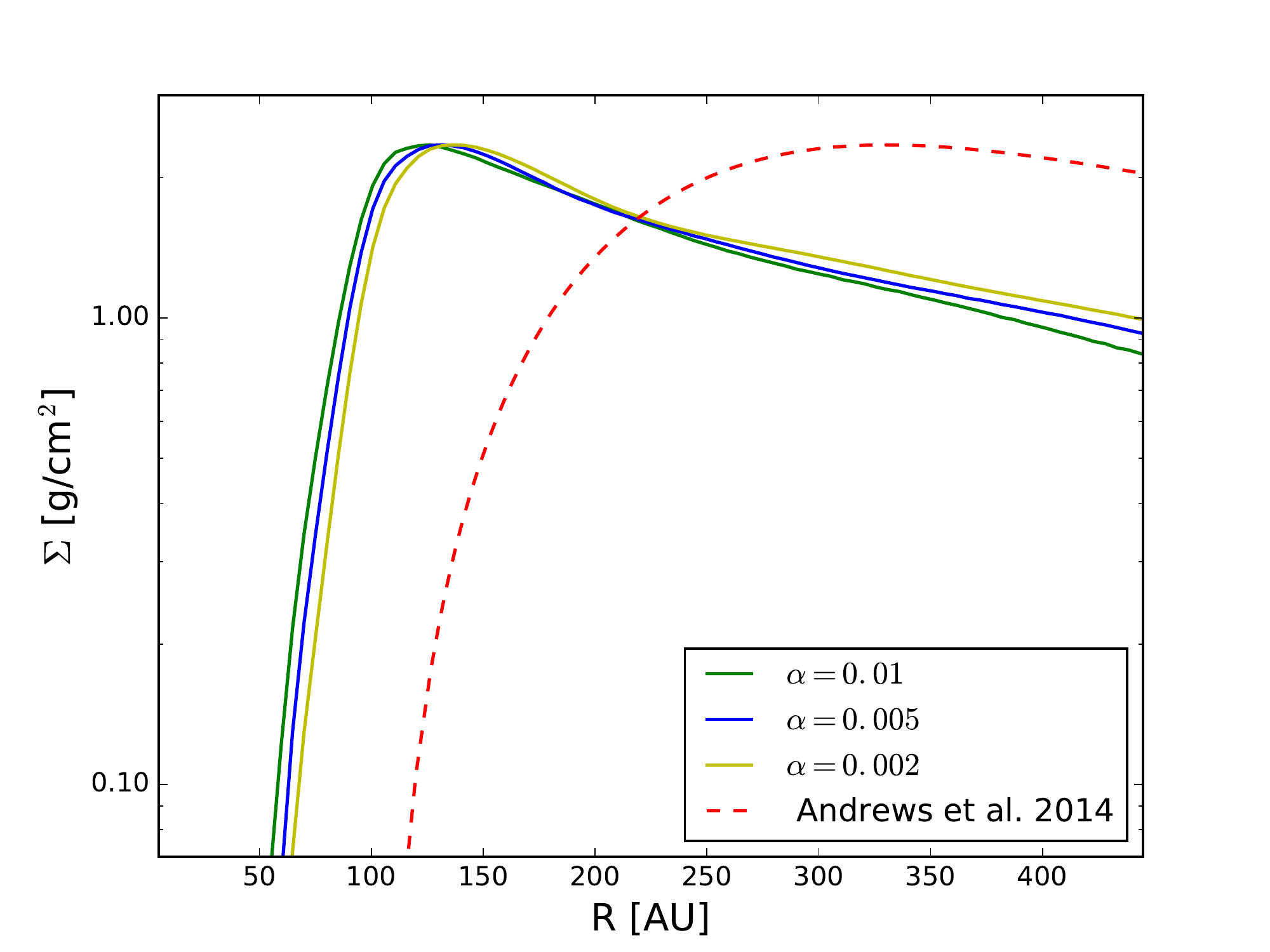}
\caption{The solid lines show density profiles obtained from our simulations in the coplanar case using $\alpha=0.01, 0.005, 0.002$. The gas density profile invoked by \citet{andrews14} is also plotted (dashed line) for a comparison.}\label{figcoplanar} 
\end{center}
\end{figure}

It is also important to test how the gas density profile is affected by different choices for the disk viscosity. Fig. \ref{figcoplanar} shows the three gas density profiles (azimuthally and temporally averaged over a few binary orbital periods in order to remove numerical noise) corresponding to the three values of viscosity tested in our simulations. These profiles are compared to the density profile modelled in \citet{andrews14}, that was proposed to reproduce the necessary dust trapping at the location where the dust ring of $\sim$mm-sized grains was observed. 

\begin{figure}[htbp!]
\begin{center}
\includegraphics[width=\columnwidth]{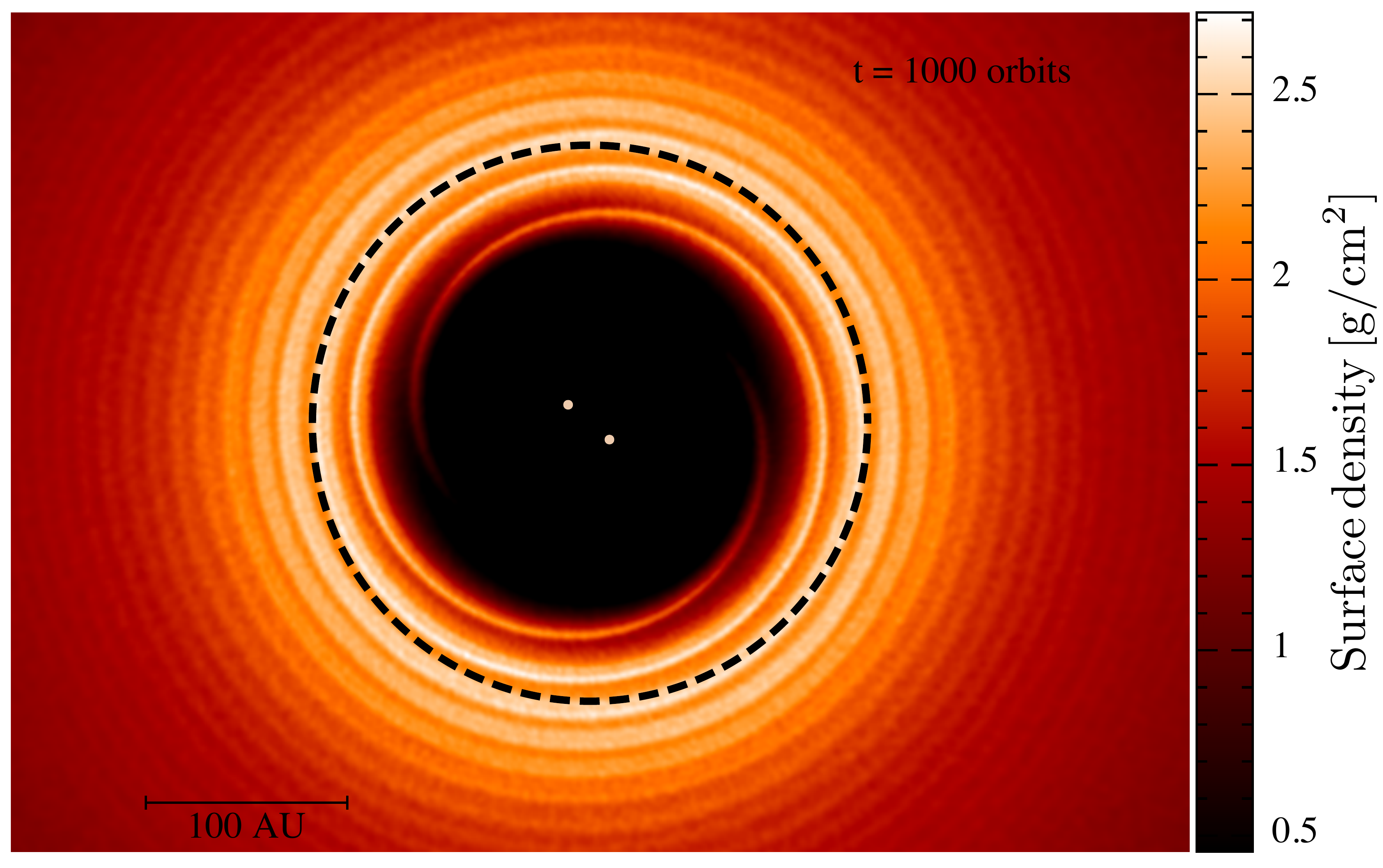}
\caption{Snapshot of the inner cavity of the circumbinary disk at 1000 binary orbits, in the $\alpha=0.002$ case. The white dots indicate the location of the two central stars, the dashed lines shows the location of the pressure maximum. The image was produced using \texttt{SPLASH} \citep{splash}.}\label{fig:sim} 
\end{center}
\end{figure}

In all the three cases, our gas simulations produce very small tidal truncation radii ($< 100\AU$), and a gas density peak located at radii of $130-140\AU$, much smaller than in the profile proposed by \citet{andrews14}. There is no strong dependence of the location of the gas density peak on the assumed value of $\alpha$. 

Fig. \ref{fig:sim} shows the inner cavity in a snapshot of our simulation $\alpha=0.002$ at 1000 binary orbits. The white dots mark the location of the two central stars, while the black dashed lines shows the location of the maximum of the gas radial density profile.

\subsection{Dust evolution}\label{sec:dust_coplanar}
We use the dust evolution model from \citet{birnstiel} to investigate the expected density distribution of dust in the disk. We apply these models to the outputs of the three \textit{coplanar case} simulations, corresponding to the three values of viscosity. 

\begin{figure}[htbp!]
\begin{center}
\includegraphics[width=\columnwidth]{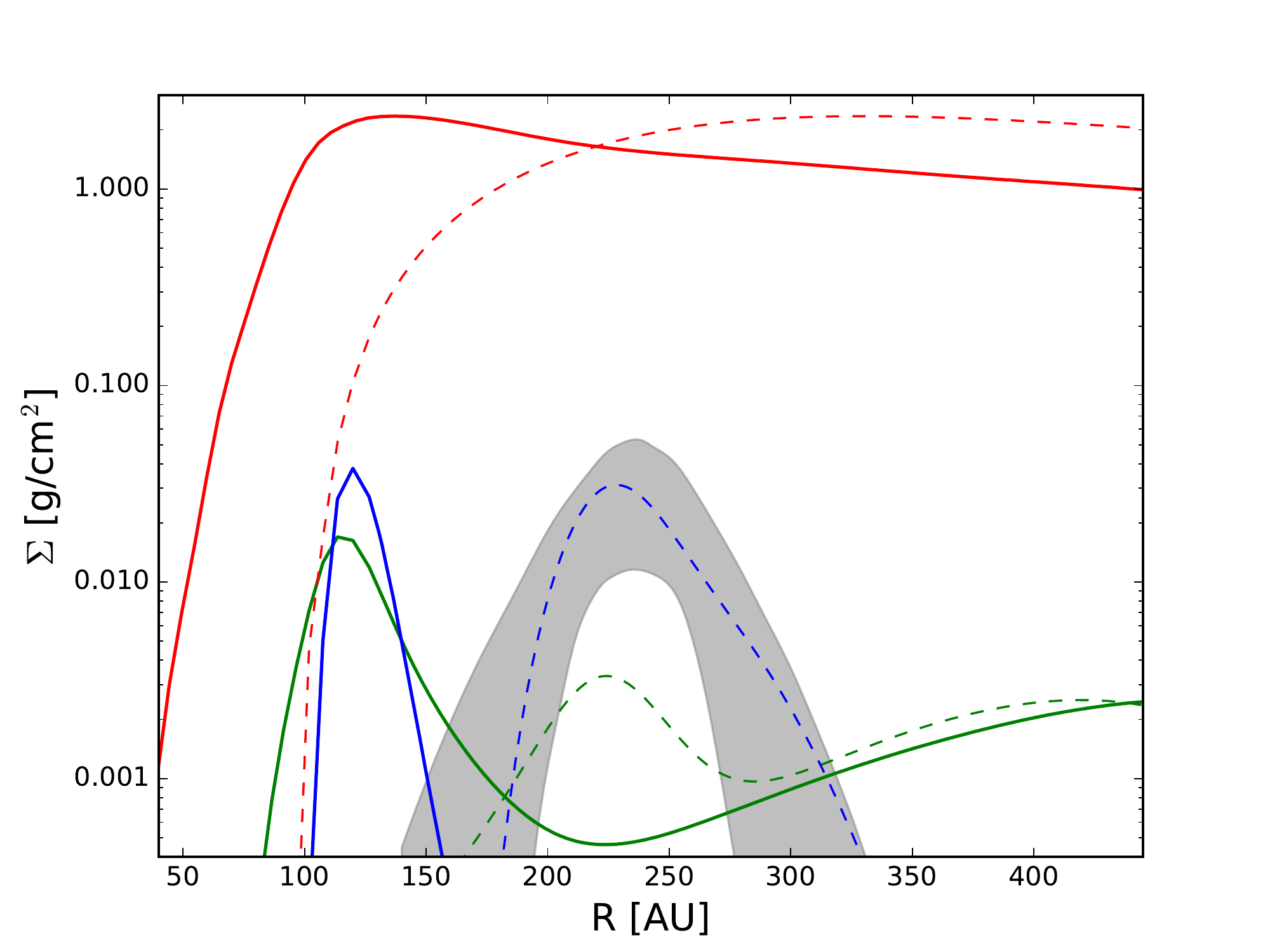}
\caption{The solid lines show the gas (in red) and dust (blue for mm-sized and green for micron-sized) density profiles obtained from our simulations in the coplanar case using $\alpha=0.002$. The gas and dust density profiles from the model by \citet{andrews14} are also plotted (dashed lines) for a comparison. It is clear that the solid-blue line showing the mm-sized dust  surface density profile is not consistent with the data (grey-shaded area).}\label{fig:til} 
\end{center}
\end{figure}

We were able to obtain dust trapping for mm-sized grains only in the case of $\alpha=0.002$. For higher $\alpha$ values, mm-sized dust particles tend to fragment to smaller sizes \citep[e.g.][]{2012A&A...539A.148B} and their trapping efficiency decreases \citep[see][]{birnstiel} . As shown in by the blue-solid line in Fig. \ref{fig:til}, at this low value of viscosity a large enough population of mm-sized grains is formed and it is efficiently trapped at $\sim 150\,\rm AU$, the location of the gas pressure maximum. However, the ring is too close to the central star and our results are inconsistent with the data, represented by the grey-shaded area in Fig. \ref{fig:til}.

This inconsistency can not be solved simply by considering different values of viscosity or a different temperature profile. In fact, as shown in section \ref{sec:viscosities} and appendix A, the gas density profile obtained from our hydrodynamical simulations does not depend strongly on these properties of the disk. Since the dust ring in mm-sized particles around GGTau A is due to dust trapping at the location of the gas pressure maximum, we can therefore conclude that the orbit of the binary and the disk cannot be coplanar: the large radial location of the ring cannot be explained by such a configuration, since the gas density maximum, and consequently the dust ring, would lie at a radius which is too small compared to the position of the ring of mm-sized dust.
The inner radius of micron sized grains is also strongly underestimated with respect to the $180-190\,\rm AU$  observed by \citet{duchene} and shown in Fig. \ref{fig:obs}.

\section{Discussion}\label{secdiscggtau}
We tested the hypothesis that the narrow dust ring around GG Tau A can be explained by dust trapping at the gas density maximum. We showed that this scenario cannot be explained consistently with astrometric measurements by a binary orbiting on the same plane of the circumbinary disk. Indeed, the best-fit orbit calculated by \citet{kohler} without fixing any parameter gives a misalignment between the disk and the orbital plane of $\sim30\degree$  However, the uncertainties on the fitted orbital parameters in this case are much larger, given the poor sampling of the binary orbit.

If, on the one hand, the resulting separation between the two stars in this coplanar case is too small to explain the location of the dust ring, on the other hand, a binary orbit with a larger semi-major axis could in principle be able to create such a wide ring. As $a$ increases, the truncation radius for the gas component of the disk moves further out, and the gas pressure bump trapping the dust is located at larger radii. The astrometric measurements for the proper motion of GG Tau A are consistent with wider binaries if the hypothesis of a disk coplanar with the binary motion is dropped and the binary and the disk are misaligned. Tab. \ref{tabkoehler} shows that orbits with $a=60-80\,\AU$ are consistent with misalignments between $25\degree$ and $30\degree$. In the latter scenario, we expect the disk to become eccentric and warped, and our approach, which assumes azimuthal symmetry, would not be suitable to test it. 3D hydrodynamical simulations including gas and dust would be instead required \citep{2014MNRAS.440.2136L,2015MNRAS.453L..73D,2016MNRAS.459L...1D}.

Tidal truncation itself is influenced by the misalignment between orbital and disk plane. Recently \citet{nixon} studied the dependence of the tidal torque on the misalignment angle at the 2:1 inner Lindblad resonance, in the case of a nearly circular disk rotating around a circular-orbit binary. Furthermore, \citet{mirandalai} quantitatively computed how the tidal truncation radius changes in misaligned systems with respect to coplanar ones, adopting a truncation criterion determined by the balance between resonant torque (which they analytically calculated for a misaligned system) and viscous torque. The latter study also took eccentric binaries into consideration. The common conclusion is that in general the torques in misaligned systems are weaker, and that circumbinary disks in such systems tend to have smaller inner radii than in the aligned case. In principle, one could therefore expect for circumbinary disks a wide range of truncation radii, and not only the classical  $~2-3a$ prediction from \citet{al}. In practice, \citet{mirandalai} show that this happens only for very misaligned systems ($\Delta i\gtrsim90\degree$): in these cases, the inner radius of a circumbinary disk can decrease down to $1-1.5a$. We conclude that for $a=60-80\,\AU$ and for the relative binary-disk misalignment, we should expect tidal torques to truncate the disk between $180$ and $240\,\AU$.

Theoretical studies showed how it is not rare that binaries and disks form with different axes of rotation. For example, \citet{1992ApJ...400..579B} showed that in the case of an elongated cloud with a rotation axis is oriented arbitrarily with respect to the cloud axis,  the disk plane (reflecting the angular momentum of the core) and the orbital plane (reflecting the symmetry of the initial core) can indeed be misaligned. Similarly, \citet{2010MNRAS.401.1505B} showed that during the star formation process, the variability of the angular momentum of the accreting material and dynamical interactions between stars can produce significant misalignment between the stellar rotation axis and the disk spin axis.  However, tidal torques tend to realign the two planes. \citet{2014MNRAS.445.1731F} calculated this alignment torque, and concluded that we should expect circumbinary disks around close (sub-AU) binaries to be highly aligned, while disks and planets around wider binaries could still be misaligned. The latter is the case of GG Tau A, where we expect $a\approx70\,\rm AU$. 

Some observations of misaligned circumbinary disks already exist. Imaging of circumbinary debris disks shows that the disk plane and the orbital plane are misaligned for some systems, such as 99 Herculis, where the mutual inclination is $\Delta i\gtrsim30\degree$ \citep{2012MNRAS.421.2264K}. Moreover, the pre-main sequence binary KH 15D is surrounded by a circumbinary disk inclined of $10\degree -20\degree$ with respect to the orbital plane \citep[e.g.][]{2004ApJ...607..913C,2013MNRAS.433.2157L}, and the FS Tau circumbinary disk appears to be misaligned with the circumstellar disks \citep{2011PASJ...63..543H}. Finally, evidence of some misalignment between the plane of the disk and the binary orbit has been found also for the HD142527 disk \citep{2013Natur.493..191C}.

\begin{figure}[htbp!]
\begin{center}
\includegraphics[width=\columnwidth]{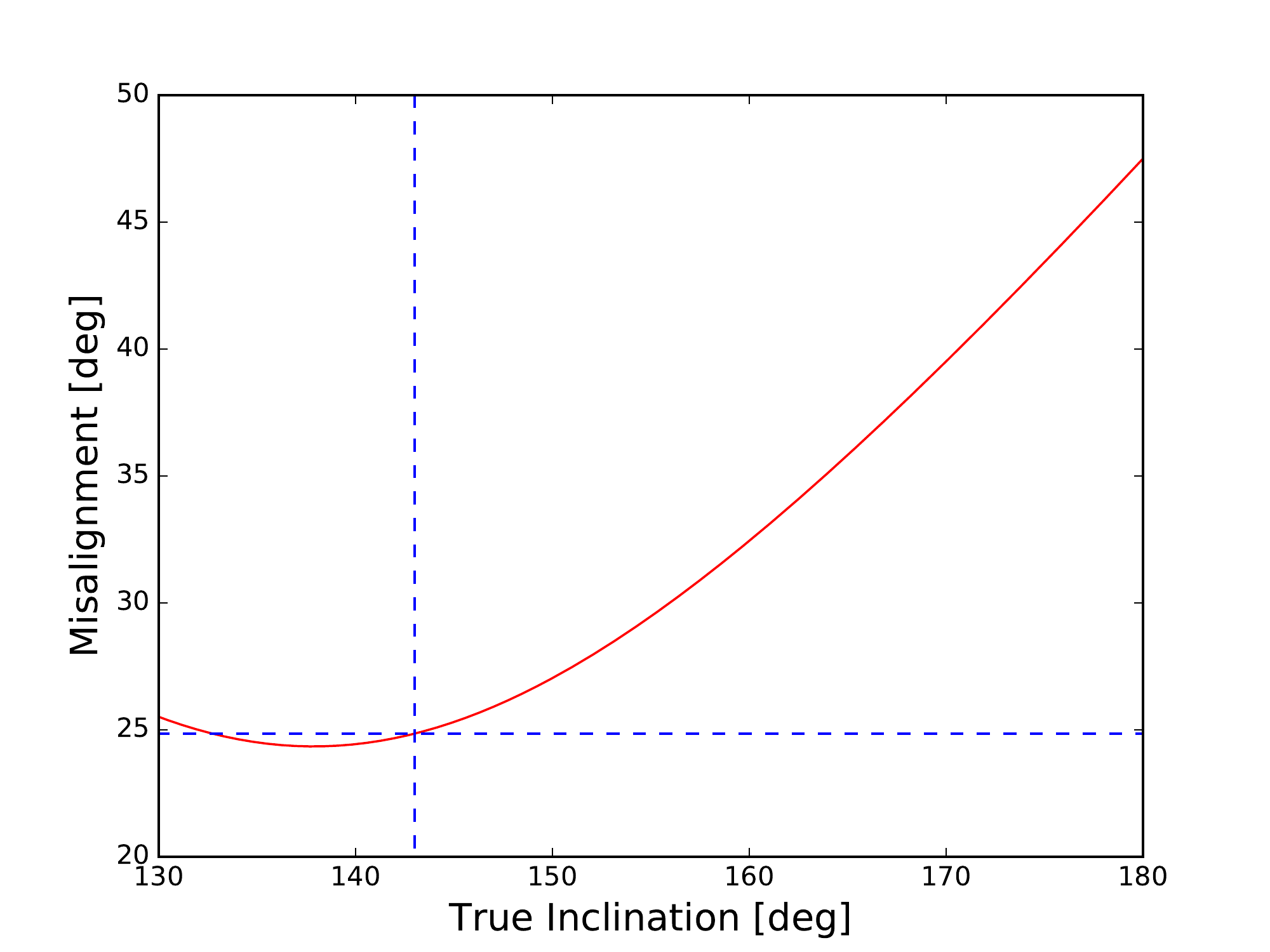}
\caption{The red line shows the misalignment $\Delta i$ between the disk and the binary orbit as a function of the disk inclination, obtained from Eq. \ref{eq:mis} for the case of GG Tau A by fixing $i_{\rm o}=132.5\degree$,  $\Omega_{\rm o}=131\degree$  and $\Omega_{\rm d}=277\degree$. The dashed blue lines mark the values of inclination calculated by assuming the disk to be circular ($i_{\rm d}=143\degree$) and the relative misalignment ($\Delta i=24.9\degree$).}\label{fig:misalignment} 
\end{center}
\end{figure}

Finally, another interesting point has recently been raised by \citet{nelson}: the inclination of the GG Tau A circumbinary disk has been calculated by assuming the disk to be circular. If this assumption is dropped, the disk inclination could be different from the commonly assumed $\sim143\degree$. Their conclusion is that it is possible to have an orbit with $a\approx60\,\rm AU$ coplanar with the disk. However, it is important to underline that a wrong estimate of the disk inclination is not enough to avoid misalignment between disk and binary orbit in the $a=60-80\,\rm AU$ cases. In a 3-dimensional space, the  angle $\Delta i$ between orbit and disk depends both on the inclinations $i_{\rm d}$ and $i_{\rm o}$ of the disk and the orbit with respect to the plane of the sky, and on the position angles of the ascending nodes $\Omega_{\rm d}$ and $\Omega_{\rm o}$ through the relation
\begin{equation}\label{eq:mis}
\cos(\Delta i)=\cos(i_{\rm o})\cos(i_{\rm d})+\sin(i_{\rm o})\sin(i_{\rm d})\cos(\Omega_{\rm o}-\Omega_{\rm d}).
\end{equation}
$\Omega_{\rm d}$ is well constrained by the gas kinematics \citep[e.g.][]{2016ApJ...820...19T}, is equal to $\sim277\degree$ and does not depend on the disk eccentricity. $i_{\rm o}$ and $\Omega_{\rm o}$ are set by the astrometric measurements and, in the case of $a=60\,\rm AU$ in Tab. \ref{tabkoehler}, we have $i_{\rm o}=132.5\degree$ and $\Omega_{\rm o}=131\degree$. Fixing these three parameters, the value $\Delta i$ calculated from Eq. \ref{eq:mis} is a function of the disk inclination alone. If the disk is eccentric, and not circular as usually assumed, then the actual disk inclination is larger than $143\degree$, and can be as big as $180\degree$. Fixing $i_{\rm o}$, $\Omega_{\rm o}$ and $\Omega_{\rm d}$ to the values above and varying the value of $i_{\rm o}$ between $143\degree$ and $180\degree$, we obtain the red curve in Fig. \ref{fig:misalignment} which clearly shows how some misalignment is always present  for all values of $i_{\rm d}$ and that it is always $>20\degree$.

In the future, observations of the proper motion of young binary systems together with high resolution observations will allow us to better study the dynamical state of these systems. In particular, we expect in the next years to have better constraints on the orbit of GG Tau A and to be able to verify the results of our work. We also expect warps to form as a consequence of the misalignment between binary and disk \citep[e.g. ][]{2014MNRAS.442.3700F}: future gas emission observations with high enough signal to noise ratio should be able to verify whether or not GG Tau A shows evidences of a warped disk. Some azimuthal asymmetry in GG Tau has already been detected in the gas emission by \cite{dutrey14} and \cite{2016ApJ...820...19T}.

\section*{Appendix  A: dependence of the final gas density profile on the temperature profile}\label{sec:tempdep}
\begin{figure}[htbp!]
\begin{center}
\includegraphics[width=\columnwidth]{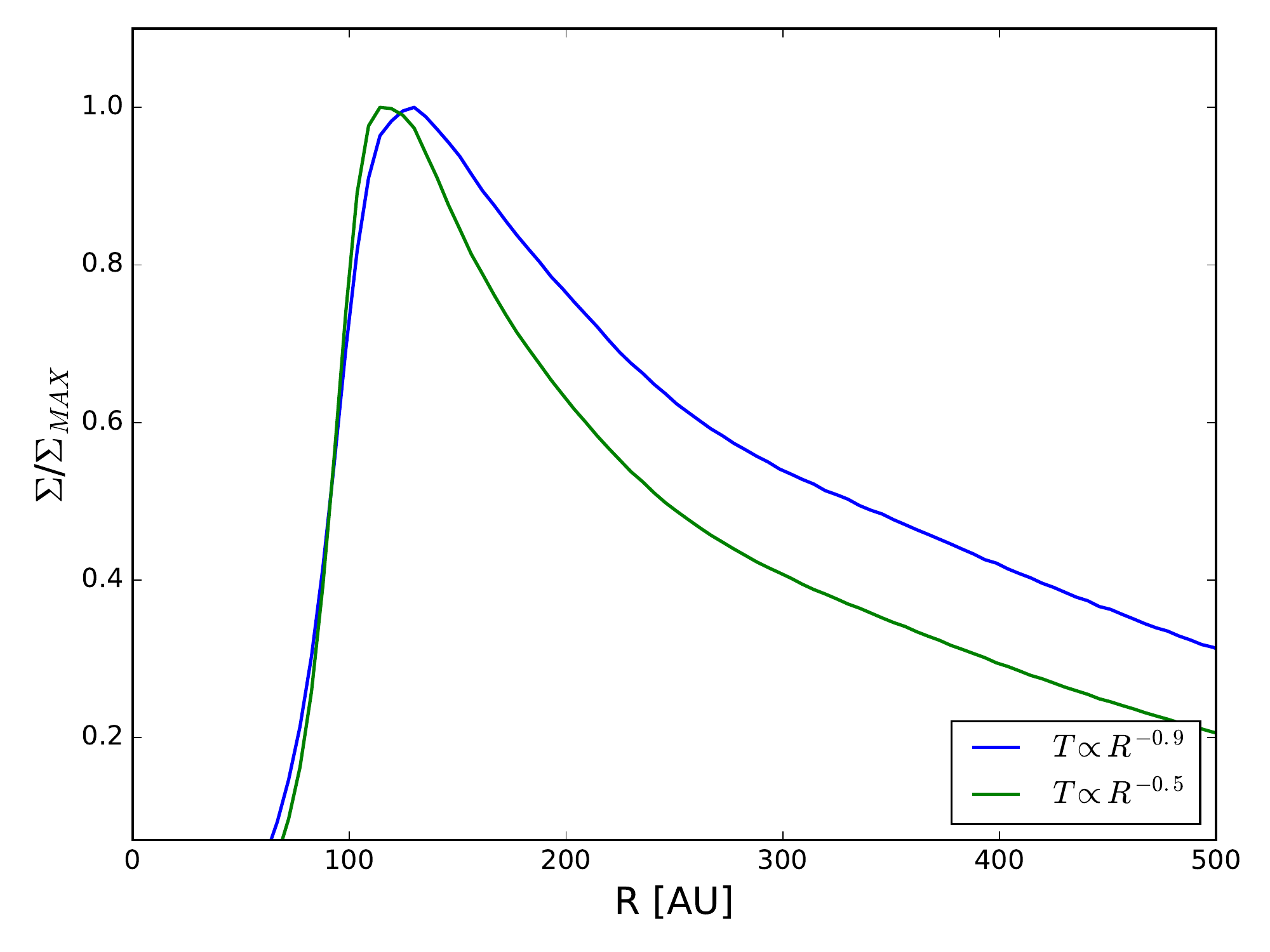}
\caption{Azimuthally and temporally averaged  gas radial density profiles,  obtained using $\alpha=0.01$ and two different temperature profiles. As expected, the density profile resulting from the $T\propto R^{-0.5}$ is steeper than the $T\propto R^{-0.9}$ one, and the density maximum in the first case is even further away from the observed dust location ($\sim 200$ AU) than in the latter case. Both the density profiles are normalised to their maximum value.}\label{fig:tprof} 
\end{center}
\end{figure}

The temperature profile in Eq. \ref{eq:tprof} is very steep. We therefore also check how the assumed temperature profile affects the steady state gas density profile and the location of the dust trap. For the coplanar case, we compare the density profiles resulting from the hydrodynamical simulations assuming the temperature profile calculated by \citet{guillo99} ($T\propto R^{-0.9}$) and a less steep and more common $T\propto R^{-0.5}$ profile in the case of $\alpha=0.01$. The two profiles are shown in Fig. \ref{fig:tprof}. In particular, we assume the temperature profile 
\begin{equation}\label{eq:tprof2}
T(R)=20\,\rm K\bigg(\frac{R}{300\,\rm AU}\bigg)^{-0.5},
\end{equation}
where the temperature at 300 AU is fixed to 20 K, as in \citet{guillo99}.

$T\propto R^{-0.5}$ leads to a steeper gas density profile, and the location of the density peak is located at smaller radii with respect to the $T\propto R^{-0.9}$  case. This shows that a less steep temperature profile does not cause an increase in the radius of the ring and that, even in the hypothesis that $T\propto R^{-0.5}$, a misalignment between the plane of the disk and that of the orbit is needed in order to explain the location of the mm-sized dust ring.

\begin{acknowledgements}
      We kindly thank R. K\"ohler for providing us with the additional orbits fitted from the astrometric measurements. We also than S. Facchini and G. Dipierro for the useful discussion, and D. Price for the support with \texttt{PHANTOM}.
\end{acknowledgements}

%-------------------------------------------------------------------

\bibliographystyle{aa}
\bibliography{paolo}

\end{document}